\documentclass{jfm}

\usepackage[pdftitle={Yongxiang Huang: Prepared for JFM},bookmarks=true,citecolor=red,colorlinks=false]{hyperref}
\usepackage{graphicx}
\usepackage{natbib}

\ifCUPmtlplainloaded \else
  \checkfont{eurm10}
  \iffontfound
    \IfFileExists{upmath.sty}
      {\typeout{^^JFound AMS Euler Roman fonts on the system,
                   using the 'upmath' package.^^J}%
       \usepackage{upmath}}
      {\typeout{^^JFound AMS Euler Roman fonts on the system, but you
                   dont seem to have the}%
       \typeout{'upmath' package installed. JFM.cls can take advantage
                 of these fonts,^^Jif you use 'upmath' package.^^J}%
      }
  \else
  \fi
\fi

\ifCUPmtlplainloaded \else
  \checkfont{msam10}
  \iffontfound
    \IfFileExists{amssymb.sty}
      {\typeout{^^JFound AMS Symbol fonts on the system, using the
                'amssymb' package.^^J}%
       \usepackage{amssymb}%
       \let\le=\leqslant  
       \let\ge=\geqslant  
      }{}
  \fi
\fi

\ifCUPmtlplainloaded \else
  \IfFileExists{amsbsy.sty}
    {\typeout{^^JFound the 'amsbsy' package on the system, using it.^^J}%
     \usepackage{amsbsy}}
    {}
\fi

\usepackage{wasysym}
\newcommand{\upd}{\mathrm{\,d}}

\newsavebox{\astrutbox}
\sbox{\astrutbox}{\rule[-5pt]{0pt}{20pt}}

\usepackage{xcolor}
\usepackage{lineno}

\setpagewiselinenumbers
\modulolinenumbers[1]
\linenumbers

\title{ Lagrangian Cascade  in Three-Dimensional Homogeneous and Isotropic Turbulence}

\author[Y.X. Huang \& F.G. Schmitt]%
{Yongxiang HUANG$^{1}$\thanks{Email address for correspondence: yongxianghuang@gmail.com} \ns 
Fran\c{c}ois G. SCHMITT$^{2,3,4}$\break
}

 \affiliation{$^1$Shanghai Institute of Applied Mathematics and Mechanics, and Shanghai Key Laboratory of Mechanics in Energy Engineering,  Shanghai University,
Shanghai 200072, PR  China\\
$^2$Universit\'e Lille Nord de France, F-59000 Lille, France\\
$^3$USTL, LOG, F-62930 Wimereux, France\\
$^4$CNRS, UMR 8187, F-62930 Wimereux, France\\
}

\pubyear{2013} \volume{xxx}
\pagerange{xx}
\date{xxx}
\begin{document}
\linenumbers
\maketitle

\begin{abstract}
In this  work, the scaling statistics of the dissipation along Lagrangian trajectories are investigated
  by using  fluid tracer particles obtained 
from a high resolution direct numerical simulation with $Re_{\lambda}=400$.  Both 
the energy dissipation rate $\epsilon$ and the local time averaged  
$\epsilon_{\tau}$ agree  rather well with the lognormal distribution hypothesis. Several statistics are then examined. It is found that the  autocorrelation function $\rho(\tau)$ of $\ln(\epsilon(t))$ and
variance $\sigma^2(\tau)$ of  $\ln(\epsilon_{\tau}(t))$ obey a log-law   with  scaling exponent $\beta\rq{}=\beta=0.30$ compatible with the intermittency parameter $\mu=0.30$. The $q$th-order moment of $\epsilon_{\tau}$ has a clear power-law on the inertial range $10<\tau/\tau_{\eta}<100$. The measured scaling exponent $K_L(q)$ agrees remarkably with $q-\zeta_L(2q)$ where $\zeta_L(2q)$ is the scaling exponent  estimated using the Hilbert methodology.  All these results suggest that the dissipation along Lagrangian trajectories could be  modelled by  a  multiplicative cascade. 
\end{abstract}

\section{Introduction}
  Turbulent flows are complex and multiscale and are characterized by eddy motions of different spatial sizes  with different time scales
  \citep{Frisch1995,Pope2000,Tsinober2009book}. This has been described by  Kolmogorov\rq{}s scaling theory of turbulence in 1941.
The scaling 
behavior of the Eulerian velocity field has been studied in details to quantify the 
intermittent 
nature of turbulence for large Reynolds numbers and homogeneous turbulence  \citep{Sreenivasan1997,Frisch1995}. Here we consider the 
Lagrangian frame, in which the fluid particle is tracked experimentally or 
numerically \citep{Mordant2002PRL,Yeung2002ARFM,Chevillard2003PRL,Biferale2004PRL,Xu2006PRL,Chevillard2006PRL,Toschi2009ARFM,Meneveau2011ARFM}. 
Fluid particles are  fluctuating over different time scales with a power-law behavior in the inertial range, e.g., $\tau_{\eta}\ll\tau\ll T_{L}$, in which $\tau_{\eta}$ is the Kolmogorov time scale, and $T_{L}$ is the Lagrangian integral time scale. Conventionally, multiscale statistics are characterized by using the Lagrangian structure-functions (LSFs), i.e.,
$
S_L^q(\tau)=\langle \vert \Delta_{\tau}u(t)\vert^q\rangle\sim \epsilon^{q/2} \tau^{q/2}
$
in which $\Delta_{\tau}u(t)=u(t+\tau)-u(t)$ is the Lagrangian velocity increment, and 
$\tau$ is the separation time scale and is lying in the inertial range. Recently, 
\citet{Huang2013PRE}  showed a clear Lagrangian inertial range on the frequency range $0.01<\omega\tau_{\eta}<0.1$ (resp. $10<\tau/\tau_{\eta}<100$)  and  retrieved the scaling exponent $\zeta_L(q)$ by using a 
Hilbert-based methodology. 
The measured scaling exponents $\zeta_{L}(q)$ are nonlinear and concave, showing that intermittency corrections are 
indeed relevant for Lagrangian turbulence \citep{Borgas1993Lagrangian,Chevillard2003PRL,Biferale2004PRL,Xu2006PRLb}.  Using a high-resolution Lagrangian turbulence database, we can now verify the  scaling relations associated with the 
Kolmogorov refined similarity hypothesis (RSH).

\section{Refined Similarity Hypothesis}
Let us recall some main ingredients of the Lagrangian version of the Kolmogorov\rq{}s refined similarity hypothesis (LRSH), in which the energy dissipation rate $\epsilon=2\nu S_{ij}S_{ij}$ is involved. Here $S_{ij}=1/2(\partial u_i/\partial x_j+\partial u_j/\partial x_i)$ is the velocity strain rate  tensor along a Lagrangian trajectory.  
A local time averaged of the energy dissipation rate along a Lagrangian trajectory is defined as, i.e.,
\begin{equation}
 \epsilon_{\tau}(t)=\frac{1}{\tau}\int_{0\le t\rq{}\le \tau}\epsilon(t+t\rq{})\upd t\rq{},\, X_{\tau}(t)=\ln(\epsilon_{\tau}(t))
\end{equation}
where $X_{\tau}(t)$ is the logarithm of the dissipation. Kolmogorov\rq{}s RSH (1962) has been written in the Eulerian frame. 
The Lagrangian analogy assumes that, in the inertial range, the variance of $X_{\tau}(t)$ has a logarithmic decrease, i.e.,
\begin{equation}
\sigma^2(\tau)=A-\beta\ln(\tau)\label{eq:sigma}
\end{equation}
in which $\sigma^2$ stands for the variance of $X_{\tau}$, $\beta$ is a universal constant and $A$ might depend on the flow \citep{Kolmogorov1962}. 
A $q$th-order moment of $\epsilon_{\tau}$ is expected to have a power-law behavior in the  inertial range,  i.e.,
\begin{equation}
M_q(\tau)=\langle \epsilon_{\tau}^{q} \rangle\sim \tau^{-K_L(q)}\label{eq:moments}
\end{equation}
This relation can be completed by a logarithmic decrease of the autocorrelation function of $X(t)=\ln(\epsilon(t))$ associated to multifractal cascades \citep{Arneodo1998}, i.e.,
\begin{equation}
\rho(\tau)=\langle \tilde{X}(t)\tilde{X}(t+\tau) \rangle =A\rq{}-\beta\rq{}\ln(\tau)\label{eq:covariance}
\end{equation}
in which  $\tilde{X}=X-\langle X\rangle$. 
For a lognormal cascade, we expect $\mu=K_L(2)=\beta=\beta\rq{}$, in which $\mu$ is the intermittency parameter \citep{Schmitt2003}. 
On the other hand, one expects also a power-law behavior for the $q$th-order LSF,
$S_q^L(\tau)\sim \tau^{\zeta(q)}$. 
The LRSH assumes a relationship between these two quantities, i.e.,
\begin{equation}
S_L^q(\tau)\sim \langle \epsilon_{\tau}^{q/2} \rangle \tau^{q/2}\sim \tau^{q/2-K_L(q/2)}
\end{equation}
leading to a relation between scaling exponents, i.e.,
\begin{equation}
\zeta_{L}(q)=q/2-K_L(q/2)\label{eq:LK62}
\end{equation}
Let us note that this scaling relation (Eq. \ref{eq:LK62}) can be found 
 for non-lognormal cascades so that the original Kolmogorov assumption of 
 lognormality  is not included in the RSH.
The original RSH in the Eulerian frame has been very well
verified \citep{Stolovitzky1992PRL,Stolovitzky1994,Chen1997PRL,Praskovsky1997PoF}.  
However, only few  works have tested the RSH in Lagrangian frame  \citep{Chevillard2003PRL,Yu2010PRL,Benzi2009PRE,Sawford2011PoF,Homann2011PoF}.  For example,   \citet{Chevillard2003PRL} proposed a multifractal formula to describe the Lagrangian velocity increments. It is found that the left part of the measured singularity spectrum $D(h)$  agrees  well  with both the lognormal model and the log-Poisson model.
\citet{Yu2010PRL} 
investigated the Lagrangian time correlation function $\rho(\tau)$ for both 
Lagrangian strain- and rotation-rate tensors. They found that the correlation function $\rho(\tau)$  depends on the spatial location of particles released.  \citet{Benzi2009PRE} tested the LRSH along  Lagrangian trajectories.  They showed that the LRSH is well verified by making Extended Self-Similarity plots. 
\citet{Homann2011PoF} studied a conditional Lagrangian increment statistics. They found that the intermittency is significantly  reduced when the LSF is conditioned on the energy dissipation rate or similar quantities (e.g.,  square of vorticity). A similar result has been shown for the Eulerian 
velocity structure-function: it is found that if one removes strong dissipation events, the corresponding scaling exponent is then approaching  the Kolmogorov\rq{}s 1941 ones without intermittent correction \citep{Kholmyansky2009PLA}. 
Note that in all these studies,  the relation \ref{eq:LK62} was not directly tested. 

We would like to provide a comment on the LSF.  It has been shown that due to the influence of large scale motions, known as infrared effect, and to the contamination of small scales, known as ultraviolet effect, the classical LSF mixes the large and small scales information \citep{Huang2013PRE}. Therefore, without the help of ESS, the LSF can not identify the correct scaling behavior of the Lagrangian velocity \citep{Mordant2002PRL,Xu2006PRL,Sawford2011PoF,Falkovich2012PoF}. With the help of a fully adaptive method, namely Hilbert-Huang transform,  a clear inertial range can be found \citep{Huang2013PRE}. 
In this paper, the LRSH is verified by considering Eqs.\,\ref{eq:sigma}, \ref{eq:moments}, \ref{eq:covariance} and \ref{eq:LK62}, where the scaling exponents for the Lagrangian velocity are extracted by using a Hilbert-based method.

\section{Numerical Validation}

\begin{figure}
\centering
 \includegraphics[width=0.65\linewidth,clip]{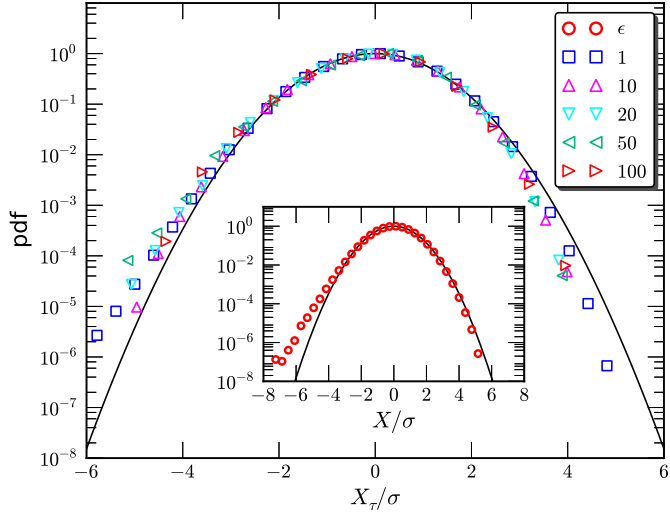}
  \caption{(Color online)  Measured pdf of the $X=\ln(\epsilon)$ and local averaged $X_{\tau}=\ln(\epsilon_{\tau})$ for several time scales in both dissipative ($\tau/\tau_{\eta}<10$) and inertial ($\tau/\tau_{\eta}\ge10$) ranges.  For comparison, the normal distribution is illustrated by a solid line. Graphically, except for   values with $\vert X_{\tau}\vert>4\sigma$,  the measured pdfs agree well with the lognormal distribution. Note that for display clarity, the measured pdf has been centered and vertically shifted by plotting $p(X_{\tau})=p_{\tau}(X_{\tau})/p_{\max}(\tau)$, in which  $p_{\max}(\tau)=\max_{X_{\tau}}\left\{p_{\tau}(X_{\tau})\right\}$.
  }\label{fig:pdf}
\end{figure}

\begin{figure}
\centering
 \includegraphics[width=0.65\linewidth,clip]{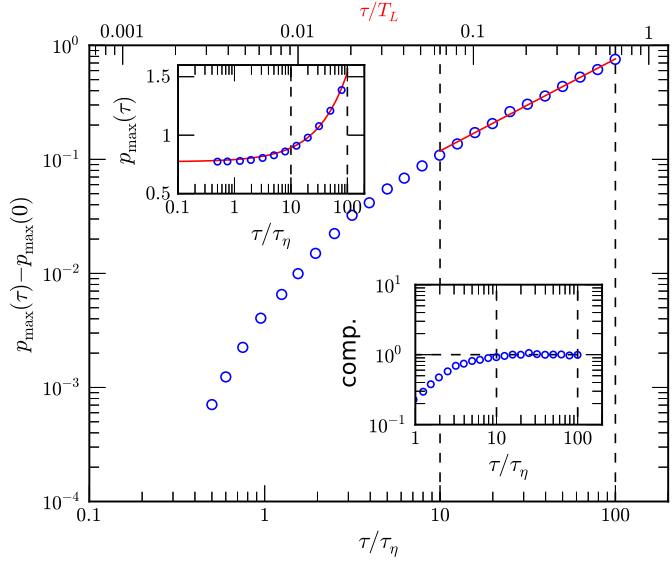}
  \caption{(Color online)  Measured max value of pdfs, $p_{\max}(\tau)-p_{\max}(0)$, in which the inertial range $10\le \tau/\tau_{\eta}\le 100$  is illustrated by a dashed line. The upper inset shows the measured $p_{\max}(\tau)$, in which the solid line is the power-law fitting. The lower inset shows the compensated 
  curve $\left(p_{\max}(\tau)-p_{\max}(0) \right)\tau^{-0.81}$. A power-law behavior is observed in the inertial range $10<\tau/\tau_{\eta}<100$ with a scaling exponent $\alpha=0.81\pm0.03$. The statistical error of $\alpha$ is the difference between the scaling exponent fitted on the first and second half of the scaling range (in log scale).
  }\label{fig:pmax}
\end{figure}

\begin{figure}
\centering
 \includegraphics[width=0.65\linewidth,clip]{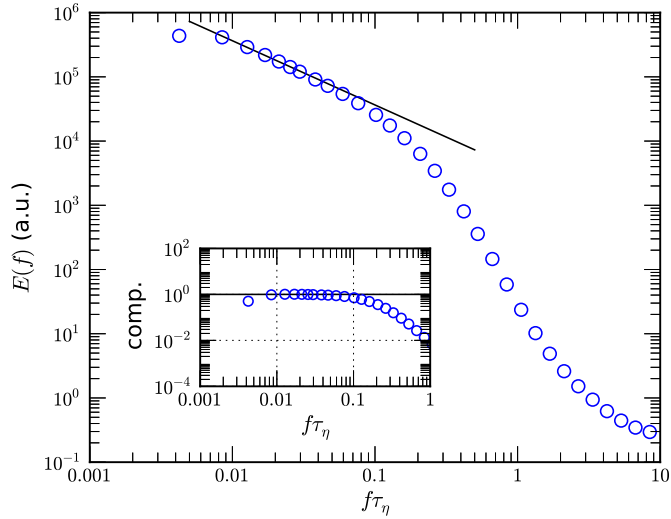}
  \caption{(Color online) Measured Fourier power spectrum of $\ln(\epsilon)$. A power law behavior is observed on the range $0.01<f\tau_{\eta}<0.06$ with a scaling exponent $1.06\pm0.13$. The inset shows the compensated curve with fitted parameters. 
  }\label{fig:psd}
\end{figure}

\begin{figure}
\centering
 \includegraphics[width=0.65\linewidth,clip]{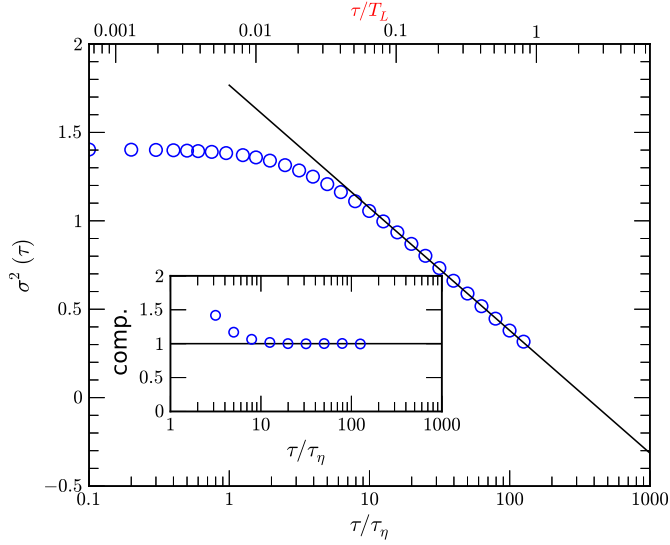}
  \caption{(Color online)  Measured  variance $\sigma^2(\tau)$ ($\ocircle$)  of $\ln(\epsilon_{\tau})$ ($\ocircle$). A log-law is observed  with a  scaling exponent  $\beta=0.30\pm0.01$ on the range $10<\tau/\tau_{\eta}<100$. The inset shows the compensated curve with fitted parameters to emphasize the log-law.  The statistical error of $\beta$ is estimated as in Fig.\,\ref{fig:pmax}.
  }\label{fig:variance}
\end{figure}

\begin{figure}
\centering
 \includegraphics[width=0.8\linewidth,clip]{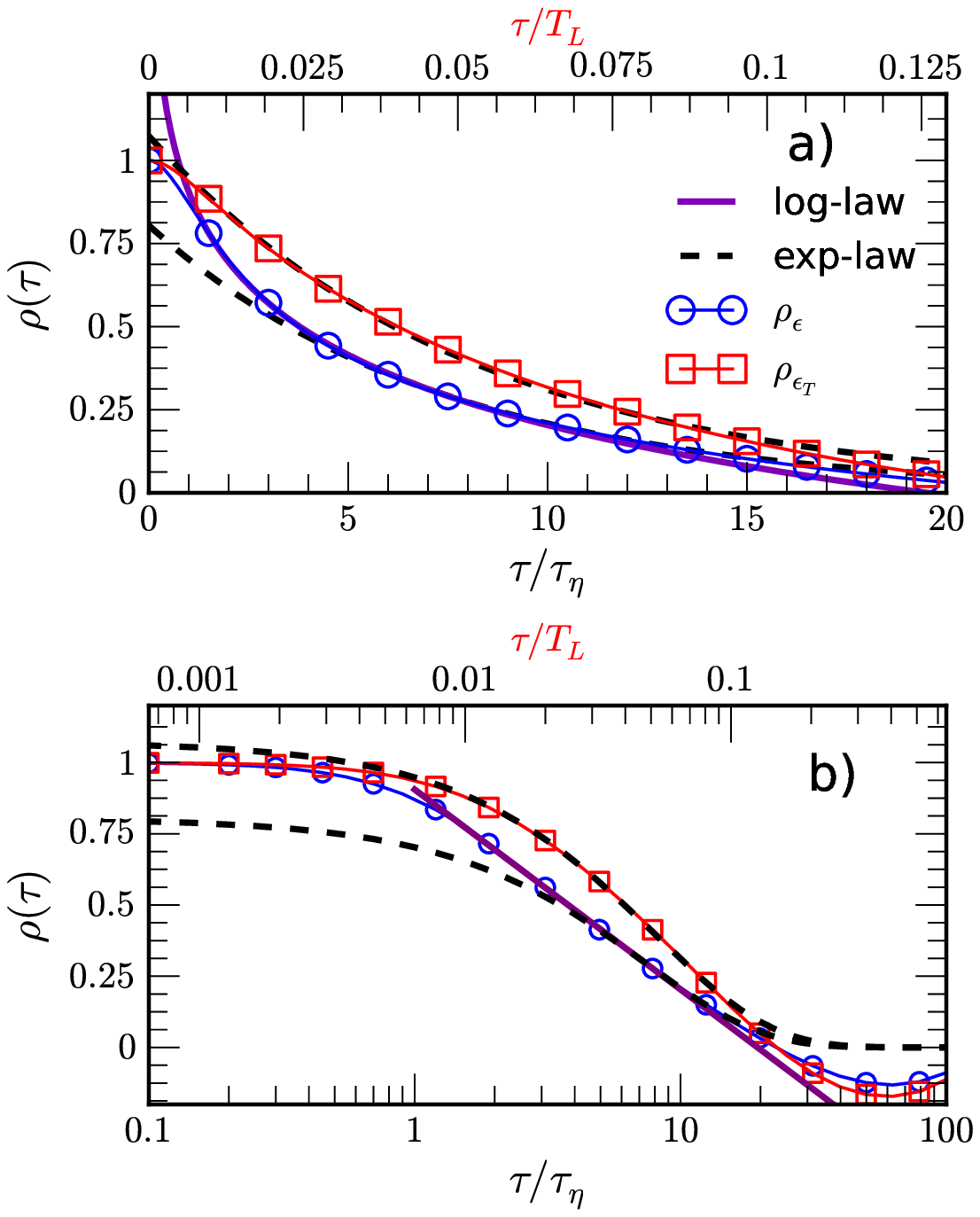}\\
  \includegraphics[width=0.8\linewidth,clip]{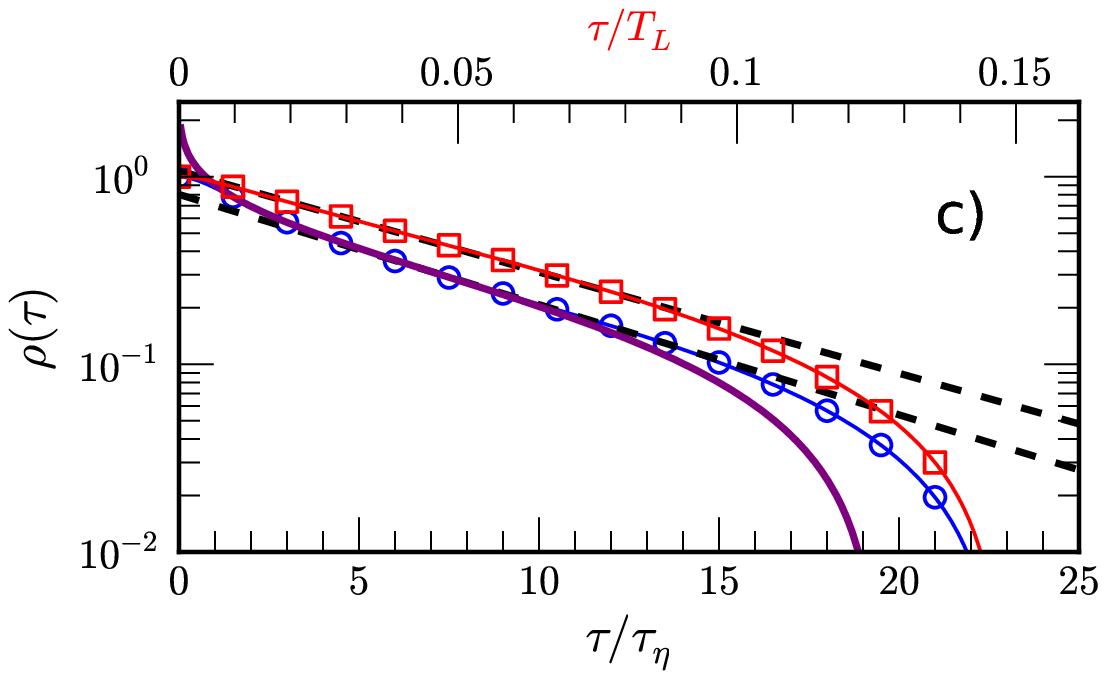}
  \caption{(Color online)  Measured autocorrelation function $\rho(\tau)$ for the logarithm of the energy dissipation rate $\epsilon$ ($\ocircle$) and pseudo-dissipation $\epsilon_T$ ($\square$). a) lin-lin plot, in which the dashed line and solid line are respectively exponential and logarithmic fitting. b) semilogx plot
   and c) semilogy plot.
 An exponential law is observed respectively on the range $3<\tau/\tau_{\eta}<15$ (resp. $0.019<\tau/T_{L}<0.097$) for $\epsilon$ and on the range $0<\tau/\tau_{\eta}<15$ (resp. $0<\tau/T_{L}<0.097$) for $\epsilon_T$. Log-law fitting is observed on the range $1<\tau/\tau_{\eta}<15$ (resp. $0.0065<\tau/T_{L}<0.097$) with a scaling exponent $\beta\rq{}=0.30\pm0.01$ for the full dissipation $\epsilon$, verifying Eq. \ref{eq:covariance}.
  }\label{fig:autocorrelation}
\end{figure}

\begin{figure}
\centering
 \includegraphics[width=0.65\linewidth,clip]{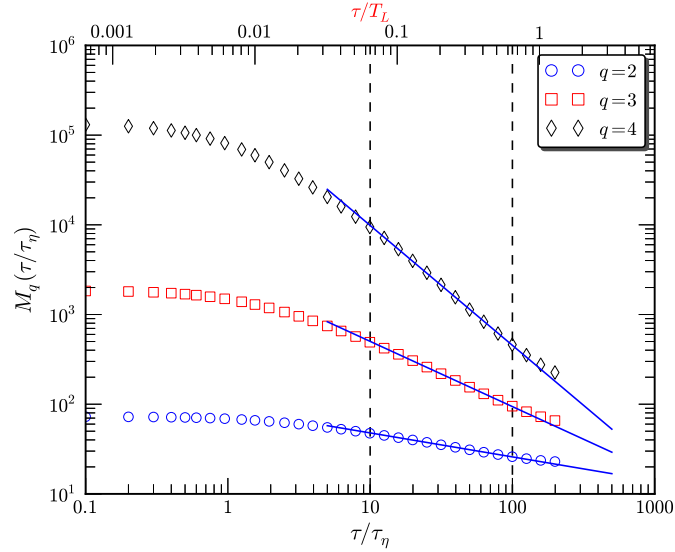}
  \caption{(Color online)  Measured $M_q(t/\tau)$ for $q=2,3$ and $4$, in which the inertial range $10\le \tau/\tau_{\eta}\le 100$ is indicated by a dashed vertical line. Power law behavior is observed on this inertial range for all moments we considered here. The scaling exponents $K_L(q)$ are then estimated on this inertial range.
  }\label{fig:Qmoment}
\end{figure}

\begin{figure}
\centering
 \includegraphics[width=0.65\linewidth,clip]{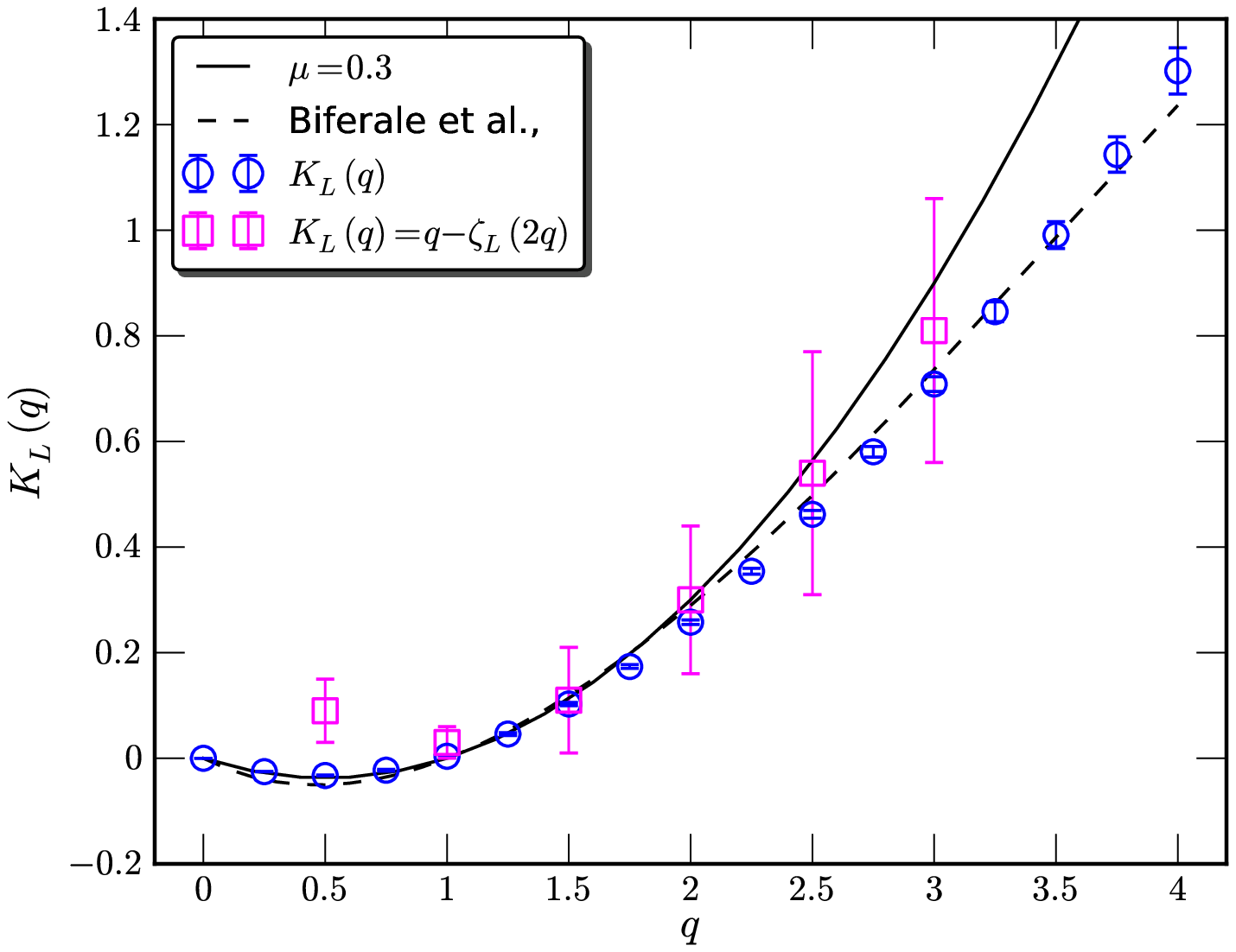}
  \caption{(Color online)  Measured $K_L(q)$ ($\ocircle$) and $q-\zeta_{L}(2q)$ ($\square$) provided by a Hilbert method.  For comparison, the curve predicted by the lognormal model with an intermittent parameter $\mu=0.30$ (solid line) and the log-Poisson based  multifractal model (dashed line) are also shown.  
   The errorbar is the difference between the scaling exponent fitted on the first and second half of the inertial range (in log scale).
  }\label{fig:Kq}
\end{figure}


The dataset considered here is composed by Lagrangian velocity trajectories in a  homogeneous and isotropic turbulent flow obtained from a $2048^3$   DNS simulation with a Reynolds number $Re_{\lambda}=400$. We recall briefly some key parameters of this database. There are $\sim 2\cdot10^5$
fluid tracer trajectories, each composed by  $N=4720$ time sampling  saved every
$0.1 \tau_{\eta}$ time units, in which $\tau_{\eta}$ is the Kolmogorov time scale.
Hence, we can access time scales in the range $0.1<\tau/\tau_{\eta}<236$.   The integral time scale $T_L$ is estimated by using the well-known Kolmogorov scaling relation as, i.e., $T_L/\tau_{\eta}=Re^{1/2}\simeq 155$, in which  $Re=\frac{3}{20}Re_{\lambda}^2$ \citep{Pope2000}.
The full energy dissipation rate $\epsilon(t)$ is retrieved from this database along the Lagrangian trajectories.  Previously, an inertial range $0.01<\omega\tau_{\eta}<0.1$ (resp. $10<\tau/\tau_{\eta}<100$ or   $0.065<\tau/T_L<0.65$) has been reported for this database by using a Hilbert-based methodology 
\citep{Huang2013PRE}. We therefore focus on this inertial range in the following 
analysis. The details of this database 
can been found in  \cite{Benzi2009PRE}.

Figure \ref{fig:pdf} shows the measured pdf $p(X)$ and  $p_{\tau}(X_{\tau})$ for time scales in the 
dissipative ($\tau/\tau_{\eta}<10$) and inertial ($\tau/\tau_{\eta}>10$) ranges.
  For display clarity, the 
measured $p_{\tau}(X_{\tau})$ have been centered and vertically shifted by taking $p(X_{\tau})=p_{\tau}(X_{\tau})/p_{\max}(\tau)$, in which 
$p_{\max}(\tau)=\max_{X_{\tau}}\left\{p_{\tau}(X_{\tau})\right\}$. For comparison, 
the Gaussian distribution is illustrated as a solid line. Graphically, the measured pdf is slightly deviating from the Gaussian distribution when $\vert X\vert >4\sigma$. This confirms that the lognormal assumption for the energy dissipation rate approximately holds also in  the Lagrangian frame at least for the central part of the pdf for $\vert X\vert<4\sigma$.

Figure \ref{fig:pmax} shows the measured maximum value of the pdf $p_{\max}({\tau}) -p_{\max}(0)$, in which the inertial range  $10<\tau/\tau_{\eta}<100$ is indicated by a dashed line. Here $p_{\max}(0)$ is for the original energy dissipation rate. A power-law behavior is observed on the inertial range, i.e.,
\begin{equation}
p_{\max}({\tau}) -p_{\max}(0)\sim \tau^{\alpha}
\end{equation}
with a measured scaling exponent $\alpha=0.81\pm0.03$. Note that the statistical error of $\alpha$ is the difference between the scaling exponent fitted on the first and second half of the inertial range (in log scale).
In the upper inset, we show the measured $p_{\max}(\tau)$, in which the solid line is the power-law fitting.  In the lower inset we show the compensated curve $
\left(p_{\max}({\tau}) -p_{\max}(0)\right)\tau^{-0.81}$ to emphasize the observed power-law. A plateau is observed in the inertial range. To our knowledge, it is the first time that this pdf scaling relation  is found. We have no interpretation presently for this relation and for the value of its exponent $\alpha$.

Figure \ref{fig:psd} shows the measured Fourier power spectrum 
$X=\ln(\epsilon)$. A power-law behavior is observed in the range 
$0.01<f\tau_{\eta}<0.06$ with a scaling exponent $1.06\pm0.13$.  The inset shows a compensated curve by using the fitted parameters to emphasize the observed power-law behavior.  
Figure \ref{fig:variance} shows the measured  variance $\sigma^2(\tau)$  of $\ln(\epsilon_{\tau}(t))$ ($\ocircle$). A log-law is observed for $\sigma^2(\tau)$ respectively on the range  $10<\tau/\tau_{\eta}<100$ with a scaling exponent $\beta=0.30\pm0.01$. 
 The 
inset shows the corresponding compensated curve to emphasize the observed 
log-law.     Eq.\,\ref{eq:sigma} is thus verified. 
Note that the $1/f$ type Fourier power spectrum is a consequence of a multiplicative cascade. Hence, this result is consistent with the logarithmic decay of the variance observed here.
 We also note that \cite{Pope1990PoF} have proposed an 
Ornstein-Uhlenbeck process for the dissipation field, which  would predict a Lorentzian (or
Cauchy) spectrum, i.e. a $f^{-2}$ decreasing. Here the observed $1/f$ spectrum does not support the Ornstein-Uhlenbeck proposal.

Figure \ref{fig:autocorrelation} shows the measured autocorrelation function $\rho(\tau)$ for both the logarithm of  the energy dissipation rate $\epsilon$ ($\ocircle$) and pseudo-dissipation $\epsilon_T=\nu \frac{\partial u_i}{\partial x_j}\frac{\partial u_i}{\partial x_j}$ ($\square$): a) lin-lin plot, b) semilogx plot
 and c) semilogy plot,
 respectively.  The  measured $\rho_{\epsilon}(\tau)$ and  $\rho_{\epsilon_{T}}(\tau)$ cross zero at $\tau/\tau_{\eta}\simeq 23$.  This is consistent with the observation in Ref. \cite{Pope1990Lagrangian}.
 We test the log-law, i.e., Eq.\,\ref{eq:covariance} first, 
 see Fig.\,\ref{fig:autocorrelation}\,b).  A log-law is observed for the dissipation $\epsilon$ on the range $1<\tau/\tau_{\eta}<15$
  ($0.0065<\tau/T_{L}<0.097)$ with a scaling exponent $\beta\rq{}=0.30\pm0.01$. However, the log-law is less pronounced for the pseudo-dissipation. The log-law is illustrated by a thick solid line in Fig.\,\ref{fig:autocorrelation}. 
  We note that \cite{Pope1990PoF}  observed an exponential decay of  $\rho(\tau)$. Figure \ref{fig:autocorrelation}\,c) shows the measured $\rho(\tau)$ in semilogy plot. An exponential law is observed respectively on the range $3<\tau/\tau_{\eta}<15$ ($0.019<\tau/T_{L}<0.097$) for $\epsilon$ and  $0<\tau/\tau_{\eta}<15$ ($0<\tau/T_{L}<0.097$) for $\epsilon_T$. The exponential law is represented by a dashed line in Fig.\,\ref{fig:autocorrelation}. Moreover, the exponential law of pseudo dissipation  is more pronounced than the one of full dissipation. 
Visually, it is
difficult to make a distinction between logarithmic  and exponential laws.
  However, we note that an exponential decay as found by \cite{Pope1990PoF} is not compatible with the intermittency framework for Lagrangian statistics, which is now well accepted \citep{Chevillard2003PRL,Biferale2004PRL}. 
Despite  the scaling range, Eq.\,\ref{eq:covariance} is verified.  
Note that the autocorrelation function can be related to the Fourier power spectrum via $\rho(\tau)=\int_0^{+\infty} E(f)\cos(2\pi f\tau)\upd f$, in which $E(f)$ is the Fourier power spectrum of $X$. Therefore, except for $f=(n+1/2)/2\tau$, $n=0,1,2,\cdots$, all Fourier 
modes contribute to $\rho(\tau)$, indicating a mixing of 
large- and small-scale information.  This could be one reason for the shift of the scaling range. A similar phenomenon is  observed for the structure-function, which could be understood as a finite size effect of the range of the power-law, known as infrared effect (large-scale motions) and ultraviolet effect (small-scale motions)   \citep{Huang2010PRE,Huang2013PRE}.

The intermittency parameter $\mu=2-\zeta_L(4)=0.30\pm0.14$ provided by the Hilbert method \citep{Huang2013PRE} is consistent with the scaling exponents $\beta$  and $\beta\rq{}$ we obtained here.  Let us note here that the
covariance log-relation was not an hypothesis of Kolmogorov, and is a relation which is different from Eq. \ref{eq:sigma}. However there are some relations between them: for a lognormal multiplicative cascade with
intermittency parameter $\mu=K_L(2)$, it can be shown that the covariance of $X$ should have a log-law with
parameter $\beta\rq{}=\mu$ \citep{Kahane1985}. Kolmogorov's hypothesis for the variance of $X_{\tau}$
is also a consequence of the cascade and its parameter is $\beta=\mu$. Here we find $\mu=0.30$ and the
values for the slopes of the covariance and variance rescaling, are fully compatible with this value of the 
intermittency parameter. We also note that the direct estimation of the intermittency
parameter from the Eulerian structure function is $\mu_E=2-\zeta(6)=0.34\pm0.03$ 
(not shown here). It is also compatible with the value we obtain for Lagrangian fluctuations.
Furthermore, when the covariance has a logarithmic decay, the Fourier power spectrum
has a $-1$ scaling,  also found here. All these results are consistent, and confirm that the dissipation in the Lagrangian frame
can be described by a multiplicative cascade.

          Figure \ref{fig:Qmoment} shows the 
measured $M_q(\tau)$ for $q=2,3$ and $4$. Power-law behavior is observed for all 
moments on the inertial range $10<\tau/\tau_{\eta}<100$. The scaling exponent $K_L(q)$ is then estimated on 
this range.  Figure \ref{fig:Kq} shows the measured $K_L(q)$ ($\ocircle$) on the 
range $0<q<4$. The errorbar is  the difference between the scaling exponent fitted 
on the first and second half of the inertial range (in log scale).  For comparison, the $K_L(q)=q-\zeta_{L}(2q)$ ($\square$) provided by the Hilbert-based methodology is also shown. 
We estimated $\zeta_L(q)$ up to $q=6$ by using the Hilbert-based method \citep{Huang2013PRE}. The corresponding $q$ for $K_L(q)$ is $3$.  The definition of the errorbar is the same as the one for $K_L(q)$. The $K_L(q)$ provided by \citet{Biferale2004PRL} log-Poisson based multifractal model and by the lognormal model with the intermittency parameter $\mu=0.30$ are respectively shown as a dashed and solid line.  For $q\le3$, all symbols collapse, showing the validity  of the scaling relation of Eq.\ref{eq:LK62} predicted by the LRSH. For $q>3$, the measured $K_L(q)$ deviates from the lognormal model since the high-order $M_q(\tau)$ corresponds the statistics of the tail of the pdf and we observed deviations from the Gaussian distribution when $X>4\sigma$.

\section{Conclusion}

In summary, the scaling statistics of the energy dissipation along the Lagrangian trajectory
 is investigated  by using  fluid tracer particles obtained 
from a high resolution direct numerical simulation with $Re_{\lambda}=400$.  Both 
the energy dissipation rate $\epsilon$ and the local time averaged  
$\epsilon_{\tau}$ agree reasonably with the lognormal distribution hypothesis. 
The measured  $p_{\max}(\tau)-p_{\max}(0)$ (maximum value of a pdf) obeys a 
power law with a scaling exponent $0.81$, a result for which we have no theoretical explanation.
Several statistics of the energy dissipation are then examined. It is found that the 
autocorrelation function $\rho(\tau)$ of $\ln(\epsilon(t)) $ and variance $\sigma_{\tau}^2$ of $\ln(\epsilon_{\tau})$ obey 
log-laws  with scaling exponents compatible with the intermittency parameter
$\mu=0.30$ as expected for multiplicative cascades.
These results show that the dissipation along Lagrangian trajectories can be  modelled by  multiplicative cascades.
The $q$th-order moment of $\epsilon_{\tau}$ has a clear power-law on the inertial 
range.  The LRSH assumptions Eqs.\,\ref{eq:sigma} and \ref{eq:moments}, and scaling relation \ref{eq:LK62} are then verified.

 \begin{acknowledgments}

  This work is sponsored by the National Natural Science Foundation of China under Grant (No. 11072139, 11032007,  11272196, 11202122 and 11332006) ,  \lq{}Pu Jiang\rq{} project of Shanghai (No. 12PJ1403500) and  the 
 Shanghai Program for Innovative Research Team in Universities. 
We thank Prof. F. Toschi for sharing his DNS  database, which  are freely available from the iCFD database and is available for download at
{{http://cfd.cineca.it}}.

\end{acknowledgments}

\bibliographystyle{jfm}

\end{document}